\documentclass[twocolumn,amsmath,amssymb,aps]{revtex4}
\usepackage{graphicx}
\usepackage{dcolumn}
\usepackage{bm}
\usepackage{color}
\newcommand{\rot}[1]{#1}
\newcommand{\blau}[1]{#1}
\newcommand{\bi}[1]{Fig.~\ref{fig:#1}}
\newcommand{\e}[1]{eq.~(\ref{eq:#1})}

\newcommand{\bra}{\langle}
\newcommand{\ket}{\rangle}
\newcommand{\mbx}{\mathbf{x}}
\newcommand{\mby}{\mathbf{y}}

\newcommand{\LL}{\mathcal{L}}

\begin{document}

\preprint{APS/number TBD}

\title{Asymptotic Phase for Stochastic Oscillators}% Force line breaks with \\

\author{Peter J.~Thomas}
% \altaffiliation[Also at the ]{Bernstein Center for Computational Neuroscience, Humboldt University, Berlin, Germany.}%Lines break automatically or can be forced with \\
\affiliation{%
Bernstein Center for Computational Neuroscience. Humboldt University, 10115 Berlin, Germany.\\
Department of Mathematics, Applied Mathematics, and Statistics.
Case Western Reserve University,
Cleveland, Ohio, 44106, USA.
}%

\author{Benjamin Lindner}
\affiliation{Bernstein Center for Computational Neuroscience and Department of Physics. 
Humboldt University, 10115 Berlin, Germany.}

\date{\today}

\begin{abstract}
Oscillations and noise are ubiquitous in physical and biological systems.  When oscillations arise from a deterministic limit cycle, entrainment and synchronization  may be analyzed in terms of the asymptotic phase function.  In the presence of noise, the asymptotic phase is no longer well defined.  We introduce a new definition of  asymptotic phase  in terms of the slowest decaying modes of the Kolmogorov backward operator.  Our \textit{stochastic asymptotic phase} is well defined for noisy oscillators, even when the oscillations are noise dependent.  {It reduces to the classical asymptotic phase in the limit of vanishing noise.}  {The phase can be obtained either by solving an eigenvalue problem, or by empirical observation of an oscillating density's approach to its steady state.}
\end{abstract}

\pacs{Valid PACS appear here}% PACS, the Physics and Astronomy
\maketitle

%\tableofcontents

%\section{Stochastic Asymptotic Phase for a Robustly Oscillatory Stochastic Differential Equation}

%Article content.

%\subsection{A Subsection}

{\sl Introduction.} Limit cycles (LC) appear in deterministic models of nonlinear oscillators such as spiking nerve cells \citep{ErmentroutTerman2010book}, central pattern generators {\cite{Ijspeert:2008:NeuralNet}}, 
and nonlinear circuits \cite{Jordan+Smith:NODE-book:2007}.
The reduction of  LC systems to one-dimensional ``phase" variables is an  indispensable tool for understanding entrainment and synchronization of weakly coupled oscillators \cite{ErmentroutKopell1984SIAMJMathAnal,PikovskyRosenblumKurths2001}. 
Within the deterministic framework, all initial points converge to the LC,
on which we can define a phase that progresses at a constant rate ($\dot{\theta}=\omega_{LC}=2\pi/T_{LC}$). The phase $\theta(\mbx_0)$ of any point $\mbx_0$ is then defined by the asymptotic convergence of the trajectory to that phase on the LC.
\rot{However, stochastic oscillations are ubiquitous, for example in biological systems} \cite{noisy-oscillator-refs}, and in this setting the classical definition of the phase breaks down. 
For a noisy dynamics,  all initial densities will converge to the same stationary density. Thus the large-$t$ asymptotic behavior no longer disambiguates initial conditions, and the classical asymptotic phase is not well defined.

Schwabedal and Pikovsky attacked this problem by defining the phase for a stochastic oscillator in terms of the mean first passage times (MFPT) between surfaces analogous to the isochrons (level curves of the phase function $\theta(\mbx)$) of deterministic LC \cite{SchwabedalPikovsky2010PRE,SchwabedalPikovsky2010EPJ,SchwabedalPikovsky2013PRL}.  %\rot{Their construction reduces to the classical phase in the case of deterministic limit cycle systems \cite{Guckenheimer1975JMathBiol}, and applies to systems with noise dependent oscillations as well. But it is defined in terms of a numerical construction procedure, the mathematical foundations for which are unclear.}  
Here we formulate an alternative definition that is tied directly to the asymptotic behavior of the density, rather than the first passage time, and is grounded in the analysis of the forward and backward operators governing the evolution of system densities. \rot{Our operator approach leads to two distinct notions of ``phase" for stochastic systems. As we argue
below, the phase associated with the backward or adjoint operator is  closely related to the classical asymptotic phase.}

{\sl General framework.} Consider the conditional density $\rho(\mby,t|\mbx,s)$, \rot{for times $t>s$,} evolving according to the forward and backward equations
%\begin{eqnarray}
%(\partial/\partial t)\rho(y,t|x,s)&=&L_y[\rho], \\(\partial/\partial s)\rho(y,t|x,s)&=&L^+_x[\rho]
%\end{eqnarray}
\begin{equation}\label{eq:density-evolution}
\frac{\partial}{\partial t}\rho(\mby,t|\mbx,s)=\LL_\mby[\rho],\,\,\, \frac{\partial}{\partial s}\rho(\mby,t|\mbx,s)=-\LL^\dagger_\mbx[\rho],
\end{equation}
where $\LL$ and $\LL^\dagger$ are adjoint with respect to the usual inner product on the space of densities. We assume that the conditional density can be written as a sum
\begin{equation}\label{eq:eigenfunction-expansion}
\rho(\mby,t|\mbx,s)=P_0(\mby)+\sum_\lambda e^{\lambda(t-s)}P_\lambda(\mby)Q^*_\lambda(\mbx),
\end{equation}
where the eigentriples $(\lambda,P,Q^*)$ satisfy
\begin{eqnarray}
\label{eq:ef}
\LL[P_\lambda]&=&\lambda P_\lambda,\,\,\,\LL^\dagger[Q^*_\lambda]=\lambda Q^*_\lambda,\\
\bra Q_\lambda | P_{\lambda'}\ket&=&\int d\mbx\,Q^*_\lambda(\mbx)P_{\lambda'}(\mbx)=\delta_{\lambda,\lambda'}.
\end{eqnarray}
Here $P_0$ is the unique stationary distribution corresponding to eigenvalue 0, $Q_0\equiv 1$, and  for all other eigenvalues $\lambda$, we assume $\Re[\lambda]<0$.  Thus, as $(t-s)\to\infty$, $\rho(\mby,t|\mbx,s)\to P_0(\mby)$.  
We refer to the system as  \emph{robustly oscillatory} if (i) the nontrivial eigenvalue with least negative real part $\lambda_1=\mu + i\omega$   is complex (with $\omega>0$), (ii) $|\omega/\mu|\gg1$ and  (iii) for all other eigenvalues $\lambda'$, $\Re[\lambda']\le 2\mu$. These conditions guarantee that the slowest decaying mode, as the density approaches its steady state, will oscillate with period $2\pi/\omega$, and decay with time constant $1/|\mu|$.  
\rot{Writing the eigenfunctions of $\lambda_1$, the slowest decaying eigenvalue of the forward and backward operators, in polar form, we have $P_{\lambda_1}=ve^{-i\phi}$ and 
 $Q^*_{\lambda_1}=ue^{i\psi}$, where $u,v\ge 0$ and $\psi,\phi\in[0,2\pi)$.}   
\blau{Asymptotically, we obtain with this notation from \e{density-evolution}
\begin{align}
\label{eq:asymp_forward}
\frac{\rho(\mby,t|\mbx,s)-P_0(\mby)}{2u(\mbx)v(\mby)} &\simeq e^{\mu (t-s)}\cos\left(\omega(t\!-\!s)+\psi(\mbx)\!-\!\phi(\mby) \right)\nonumber\\
%&=F(t-s;\mbx,\mby)
\end{align}} \rot{As we now argue, $\psi(\mbx)$, the polar angle associated with the backward eigenfunction, is the natural generalization of the deterministic asymptotic phase.}
 
\rot{For a deterministic LC system, a given asymptotic phase is assigned to points off the LC by identifying those points which \emph{at an earlier time} were positioned so that their subsequent paths would converge.  Suppose we observe a density of points $\rho(\mby,t)$ concentrated near a position on the LC corresponding to a certain phase $\theta(\mby)\approx\theta_0$. %\footnote{Note to remove later: What is the formula for   $\rho(\mbx,s)$ in the deterministic case? Is the oscillation ``obvious''?  Suppose we write the solution of the initial value problem for $d\mby/dt=A(\mby)$ with initial value $x$ at time $0$ as $y(t)=F_t(x)$, and note $x=F^{-1}_t(y)=F_{-t}(y)$. We fix the present time $t$ and location $\mby$ on or near the LC.Then $\rho_\text{earlier}(\mbx,s)=\rho(\mbx,s|\mby,t)\rho(\mby,t)_\text{now}/\rho(\mby,t|\mbx,s)=\delta(\mbx-F_{s-t}(\mby))\rho_\text{now}(\mby,t)/\delta(\mby-F_{t-s}(\mbx))$.  But this way of writing $\rho_\text{earlier}(\mbx,s)$ makes the oscillation harder to understand.}
Fixing a point $\mbx$ away from the LC, the density $\rho(\mbx,s)$ at earlier times $s<t$   will show transient oscillations with period $T_{LC}$ as the density propagates away from the stable LC in reverse time.  The oscillations observed at two distinct points $\mbx$ and $\mbx'$ will be offset by the difference in their asymptotic phase.  Looking forward in time, all trajectories will continue converging to the LC, so the density for a point away from the LC will not oscillate -- it will remain zero.}

\begin{figure}[h!]
  \centering
  \vskip1em
\includegraphics[height=7cm]{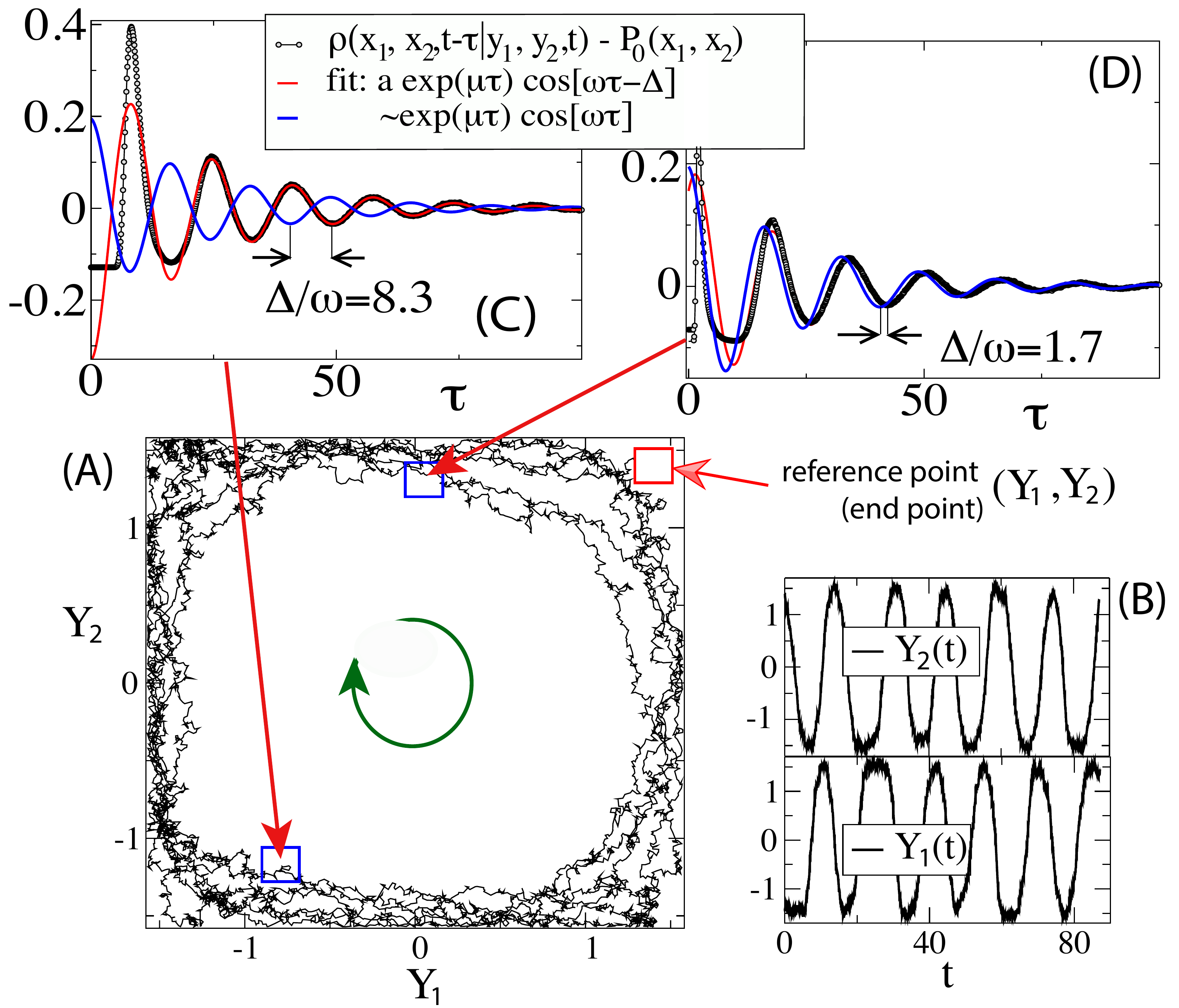}
\caption{(color online) {\bf Trajectory of the heteroclinic oscillator and the histogram method to estimate the asymptotic phase.} Trajectories in the \rot{$(Y_1,Y_2)$} plane like the one shown in (A) that all end up in the neighborhood of the reference point \rot{$(Y_1,Y_2)$} (red box) are used to estimate the time-dependent probability in the past in other points \rot{$(X_1,X_2)$} in the plane (blue boxes). This probability displays asymptotically damped oscillations (C, D), characterized by the smallest non-vanishing eigenvalue and a space-dependent phase-shift \rot{ $\Delta(x_1,x_2,y_1,y_2)=\psi(x_1,x_2)-\phi(y_1,y_2)$, from which the asymptotic phase  $\psi(x_1,x_2)$ can be extracted [the constant off-set still depends on the reference point $(y_1,y_2)$].} Stochastic oscillations of the variables are shown in (B).  
 } 
  \label{fig:histogram}
\end{figure}

\rot{Figure \ref{fig:histogram} illustrates the analogous measurement of the phase at a point $\mbx$ from the conditional density at earlier times, \blau{$\rho(\mbx,s|\mby,t)$}, for a stochastic oscillator.}
For a stationary stochastic time series this density is related to the conditional density $\rho(\mby,t|\mbx,s)$ appearing in \e{asymp_forward} by  $\rho_0(\mbx,s;\mby,t)=\rho(\mby,t|\mbx,s)P_0(\mbx)=\rho(\mbx,s|\mby,t)P_0(\mby) $ \rot{(not to be confused with the detailed balance condition)}, 
which can be used to rewrite \e{asymp_forward} as follows
\begin{equation}
\frac{\rho(\mbx,t-\tau|\mby,t)-P_0(\mbx)}{2u(\mbx)v(\mby)P_0(\mbx)} \simeq\frac{e^{\mu \tau}}{P_0(\mby)}\cos\left(\omega\tau+\psi(\mbx)\!-\!\phi(\mby) \right),
\label{eq:asymp}
\end{equation}
\blau{where we have switched to $s=t-\tau$ with $\tau>0$. If we select from a stationary ensemble the trajectories that end up at time $t$ in $\mby$, we can estimate the conditional density $\rho(\mbx,t-\tau|\mby,t)$ and the steady state $P_0(\mbx)$. Fitting then the left-hand-side of \e{asymp} to a damped cosine in $\tau$ (see \bi{histogram}), we can by virtue of  \e{asymp} infer the phase $\psi({\mbx})$ at any point $\mbx$.}
%In contrast to the deterministic oscillator, a stochastic system will converge to a  steady state distribution both forward and backward in time.
%If the system trajectories \rot{$\mathbf{Y}(t)$} show oscillatory behavior, the steady state density will be approached by a damped oscillation {at points with nonvanishing density $P_0$.} 
%\rot{For the stationary stochastic case,} the conditional density at \rot{earlier times} $s<t$  takes the asymptotic form: 
%\begin{equation}
%\label{eq:asymp}
%\frac{\rho(\mbx,s|\mby,t)-P_0(\mbx)}{2u(\mbx)v(\mby)P_0(\mbx)} \simeq\frac{e^{\mu (t-s)}}{P_0(\mby)}\cos\left(\omega(t\!-\!s)+\psi(\mbx)\!-\!\phi(\mby) \right).
%\end{equation}
%\rot{Therefore we may obtain the backward-looking asymptotic phase $\psi(\mbx)$ through stochastic simulations of an ensemble of trajectories, as illustrated in \bi{histogram}.}  

\rot{We may also obtain the backward-looking phase by solving the eigenvalue problem \e{ef} for $Q^*$.  Comparison with the deterministic case again points to the complex angle of $Q^*$ as the analog of the classical phase. For a deterministic system, $d\mbx/dt=A(\mbx)$, the conditional density $\rho(\mby,t|\mbx,s)$ obeys  \e{density-evolution} with $\mathcal{L}^\dagger_x[Q]=\sum_i A_i(\mbx)\partial Q(\mbx)/\partial x_i$.  The function $Q_1=e^{i\theta(\mbx)}$ with $u\equiv 1$ and $\psi(\mbx)\equiv\theta(\mbx)$ is an eigenfunction of $\LL^\dagger_\mbx$ with eigenvalue $\lambda=i\omega_{LC}$.  The analogous eigenfunction of the forward operator, $\mathcal{L}_\mby[P]=-\sum_i\partial(A_i(\mby)P(\mby))/\partial y_i$, is identically zero except on the LC, at which it has a delta-mass radial distribution. Thus $P_{1}$ is unsuitable for defining a ``phase" anywhere except on the limit cycle itself.}

 {\sl Noisy Heteroclinic Oscillator.}
Consider the system 
\rot{
\begin{eqnarray}\nonumber
\dot{Y_1}&=&\cos(Y_1)\sin(Y_2) +\alpha\sin(2Y_1) + \sqrt{2D} \xi_1(t)\\
\label{eq:sde}
\dot{Y_2}&=&-\sin(Y_1)\cos(Y_2) + \alpha\sin(2Y_2) + \sqrt{2D} \xi_2(t),
\end{eqnarray}
}with $\alpha=0.1$, reflecting boundary conditions on the domain $-\pi/2\le \rot{\{Y_1,Y_2\}}\le\pi/2$, and independent white noise sources $\langle \xi_i(t) \xi_j(t') \rangle=\delta(t-t')\delta_{i,j}$.  Without noise ($D=0$) the system has an attracting heteroclinic cycle, but does not possess a finite-period limit cycle. Therefore, in the noiseless case, there is no classical asymptotic phase  \cite{ShawParkChielThomas2012SIADS}. 

For weak noise, the system displays pronounced oscillations (\bi{histogram}, B), manifest as irregular clockwise rotations in the $(\rot{y_1,y_2})$ plane (\bi{histogram}, A). We can use large trajectories and condition them on their end point (red box in 
\bi{histogram}, A). \rot{As argued above,} looking back into the past of such an ensemble of trajectories, we see for large times a damped oscillation (\bi{histogram}, C and D), the damping constant and frequency of which should be related to the real and imaginary parts of the first non-vanishing eigenvalue. Indeed, 
we have checked by fitting a damped cosine according to \e{asymp} to the counting histograms of the backward probability at different positions, that the estimate of $\mu$ and $\omega$ is largely independent of location (not shown). More importantly, fitting a damped cosine function also provides an estimate of the asymptotic phase $\psi(x_1,x_2)$ in \e{asymp}. We verified that (up to a fixed phase shift at every point $(x_1,x_2)$) the resulting phase does not depend on the choice of the reference point $(y_1,y_2)$. 

As outlined above, the asymptotic phase is also given by the complex phase of the eigenfunction for the slowest eigenvalue of the system. For the process \e{sde}, the backward operator reads \rot{explicitly}
\rot{
\begin{eqnarray}
\LL^\dagger&=&[\cos(x_1)\sin(x_2) +\alpha\sin(2x_1)]\partial_{x_1}+D\partial_{x_1}^2\nonumber\\
&+&[-\sin(x_1)\cos(x_2) +\alpha\sin(2x_2)]\partial_{x_2} +D\partial_{x_2}^2.
\end{eqnarray}  
}
We solve the eigenvalue problem \e{ef} for the system by expanding the eigenfunctions  in a Fourier basis $Q^*_{\lambda}=\sum c_{m,n,\lambda} \rot{ e^{i(mx_1+nx_2)}}$ and computing the eigenvalues and eigenvectors of the corresponding matrix equation numerically. The  leading eigenvalues are shown in \bi{hetero}C for two different noise values. Under both noise conditions, the first nonvanishing eigenvalues form a complex conjugate pair (framed) that is well separated from the remaining eigenvalues.  As we would expect, for a lower noise level ($D=0.01125$, black filled circles) this separation is more pronounced than for a higher level  ($D=0.1$, red empty circles). 
  
\begin{figure}[t!]
  \centering
\includegraphics[width=9cm]{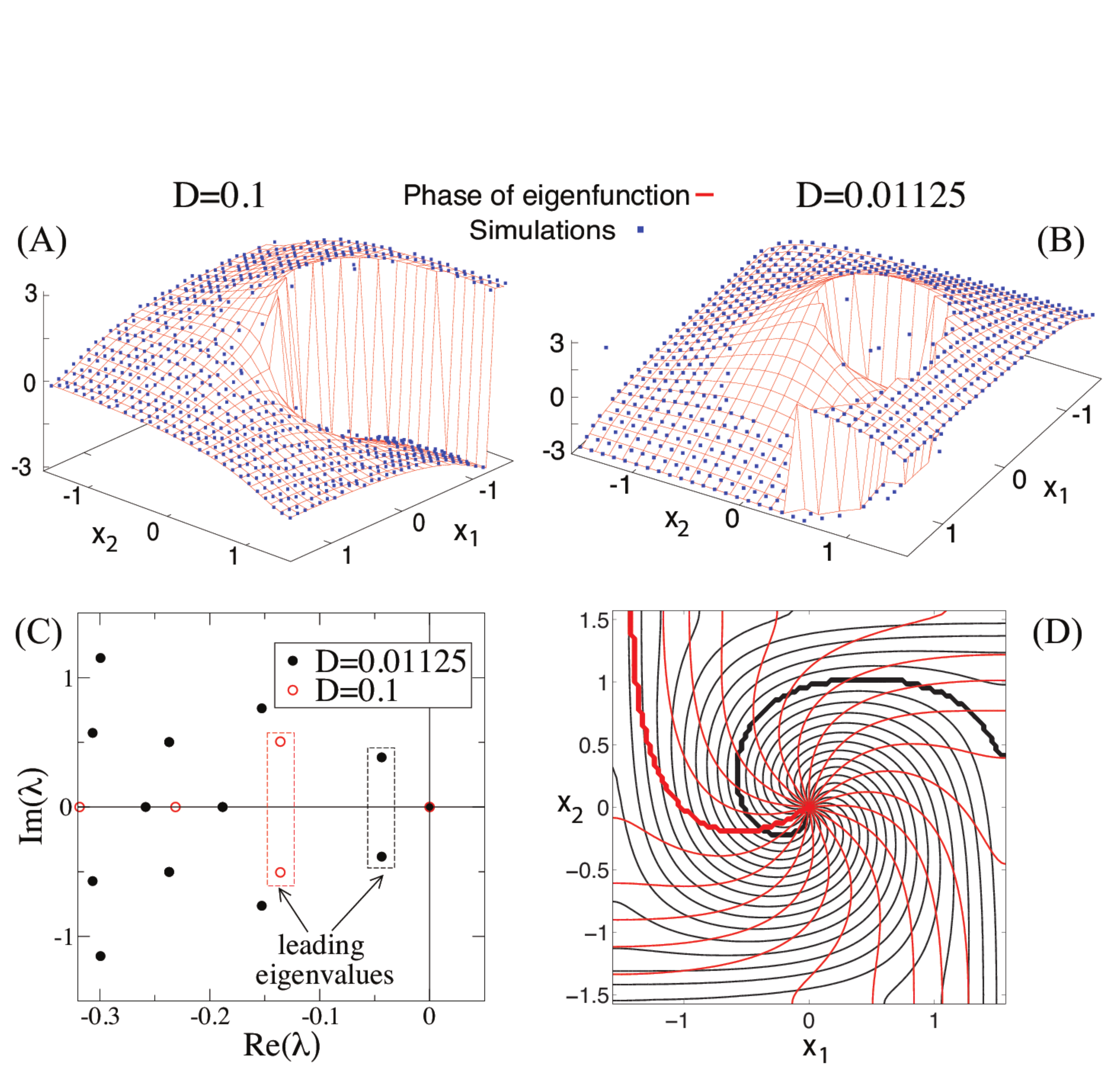}
\caption{(color online) {\bf Asymptotic phase of the stochastic heteroclinic oscillator for two different noise levels.} The complex phase of the backward eigenfunction (solid lines) is compared to the results of the histogram method \rot{\cite{comparison}} for $D=0.1$ (A) and $D=0.01125$ (B). Eigenfunctions used in (A) and (B) correspond to the slowest eigenvalues, marked by dashed boxes in (C). \rot{Isochrons} at lower noise level 
[black in (D)] are more curled than for stronger noise [red in (D)]. \rot{Thick lines in (D) denote $2\pi$-jump in phase.}} 
  \label{fig:hetero}
\end{figure}

The complex phase of the eigenfunction for the two distinct noise levels is shown in  \bi{hetero}A and B. The phase increases in the same direction as the local mean velocity (clockwise) in both cases.  For weaker noise, the phase winds inward more steeply, i.e.~the inward radial component of $\nabla\psi$ is larger. 

In  \bi{hetero}A and B we also superimpose data (blue points) generated by the histogram method, subject to a uniform constant vertical offset. The agreement of these two surfaces  demonstrates that the asymptotic phase can be obtained by the solution of the partial differential \e{ef} for model systems, for which this equation is known, but also from trajectories of the system obtained either by stochastic simulations (for a model) or measurements (experimental data). 
 
{\sl Neural Oscillator with Ion Channel Noise.} Izhikevich introduced a planar conductance-based model for excitable membrane dynamics \cite{Izhikevich2007}  that is similar to the well known two-dimensional Morris-Lecar model \cite{MorrisLecar:1981,Rinzel+Ermentrout:1989}.  We consider a  jump Markov process version of Izhikevich's model, in which noise arises from the random gating of a small, \rot{discrete} population of $N_\text{tot}$ potassium ($K$) channels, which switch between an open and a closed state.  Conditional on $N(t)$, the number of open channels at time $t$, 
the voltage $V$ evolves deterministically:
\begin{eqnarray}\nonumber
C\left.\frac{dV}{dt}\right|_N&=&I_0-I_\text{L}(V)-I_\text{NaP}(V)-I_\text{K}(V,N)\\
&=&Cf(V,N)
\label{eq:napk_v}
\end{eqnarray}
where $I_0$ is an applied current, $I_\text{L}$ is a passive leak current, $I_\text{NaP}$ is a deterministic ``persistent sodium" current and $I_\text{K}$ is a potassium current gated by the number of open potassium channels, $0\le N\le N_\text{tot}$.  We used standard parameters  \cite{details}.

The number of open channels  $N(t)$ comprises a continuous time Markov jump process with voltage dependent per capita transition rates $\alpha(v)$ for channel opening and $\beta(v)$ for channel closing \cite{Izhikevich2007}.  We generated trajectories of the joint $(V,N)$ process using an exact stochastic simulation algorithm that takes into account the time-varying transition rates $\alpha$ and $\beta$ \cite{AndersonErmentroutThomas2014JCNS,NewbyBressloffKeener2013PRL}.  \bi{napk}A shows a trajectory in the $(v,n)$ plane for $N_\text{tot}=100$ channels and applied current $I_0=60$.
The light and dark gray dashed lines show the $v$-nullcline and $n$-nullcline, respectively.  In contrast to the noisy heteroclinic oscillator, this system has a stable limit cycle in the limit of vanishing noise ($N_\text{tot}\to\infty$) with finite period $T_\text{LC}\approx 5.9825$. 
% [Where mention that the system is not well described by a Langevin equation in this range; note at $N\approx N_\text{tot}$ for a portion of each circuit of the noisy limit cycle, invalidating the Gaussian approximation for the increments distribution in a small time step.]  

The forward and backward equations for this system are \rot{given in terms of $f(v,n)$ (\e{napk_v}), $\alpha(v)$ and $\beta(v)$  \cite{details}:} 
\begin{widetext}
\begin{eqnarray}\nonumber
\frac{\partial}{\partial t}\rho(v',n',t|v,n,s)&=&\mathcal{L}_{v'}[\rho]
=-\frac{\partial}{\partial v'}\left[f(v',n')\rho \right]
-\left(\alpha(v')\left(N_\text{tot}-n'\right)+\beta(v')n'   \right)\rho\\
&&+\alpha(v')\left(N_\text{tot}-(n-1)\right)\rho(v',n'-1,t|v,n,s)+\beta(v')(n'+1)\rho(v',n'+1,t|v,n,s)\\
\nonumber
-\frac{\partial}{\partial s}\rho(v',n',t|v,n,s)&=&\mathcal{L}^\dagger_v[\rho]=f(v,n)\frac{\partial \rho}{\partial v} 
+ \alpha(v)\left(N_\text{tot}-n \right)
\left\{  \rho(v',n',t|v,n+1,s)-\rho(v',n',t|v,n,s) \right\}\\
&&+\beta(v)n\left\{ \rho(v',n',t|v,n-1,s)-\rho(v',n',t|v,n,s)\right\}
\end{eqnarray}
\end{widetext}
We approximate the operator $\mathcal{L}^\dagger$ with a finite difference scheme by discretizing
the voltage axis $-80\le v \le 20$ into 200 bins of equal width.  We obtain the eigenvalues and eigenvectors of the matrices approximating $\LL$ and $\LL^\dagger$ using standard methods (\texttt{MATLAB}, \emph{The Mathworks}).  \bi{napk}B shows the dominant (slowest decaying) part of the eigenvalue spectrum. Note the occurrence of a family of eigenvalues of the form $\lambda_k\approx \pm i\omega k - \mu k^2, k=0,1,2,\dots$.  The quadratic relationship between the real and imaginary parts of the eigenvalues of this form is consistent with the existence of a change of coordinates under which the evolution takes the approximate form of diffusion on a ring with constant drift, $\dot{\varphi}= \omega + \sqrt{2\mu} \xi(t)$.  Here the eigensystem is exactly solvable, and the spectrum lies on the same paraboloa.

In Figure 3B, the first nonzero pair (framed) for $N_\text{tot}=100$ is $\lambda_1\approx-0.031 \pm 1.0475 i$, corresponding to a period for the decaying oscillation of $T\approx5.9985$ (cf.~$T_\text{LC}$ above) and $\omega/|\mu|\approx 33.7\gg 1$. 
%\blau{Giving all these details, the numbers etc. contrasts a little with the shortness on details in my section. Reviewers may guess that these parts were written by different people ... But maybe not a main issue.} 
All other eigenvalues have real part less than or equal to $4\mu$, so the system is ``robustly oscillatory" according to our criteria (i-iii).  

\onecolumngrid
\begin{widetext}
\begin{figure}[t!]
  \centering
\includegraphics[height=4.5cm]{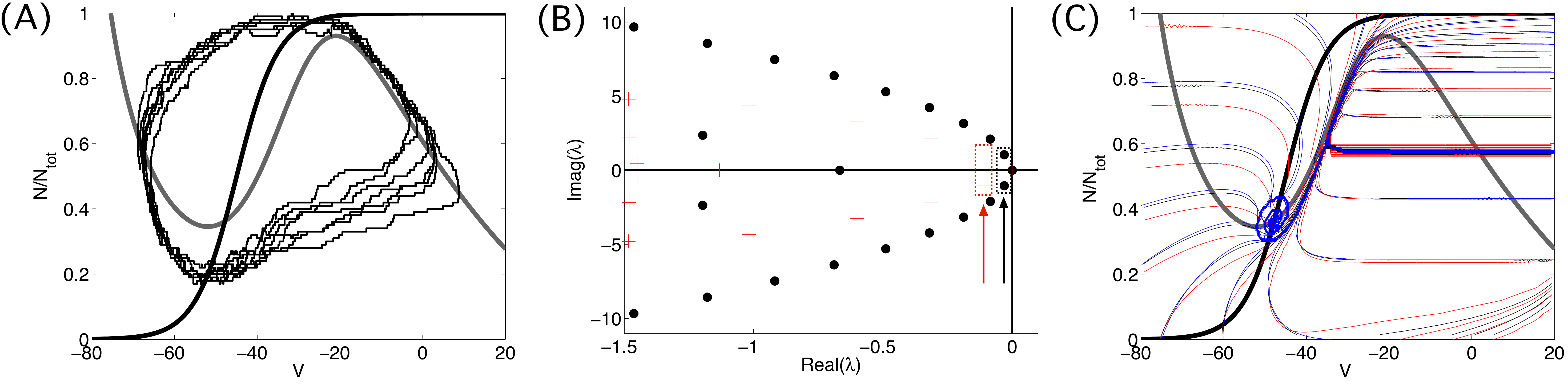}
\caption{(color online) {\bf Trajectory, nullclines, eigenvalues of the backward operator, and asymptotic phase lines for the persistent-sodium--potassium model.} (A) Sample trajectory (\rot{thin black line}) for the $(V,N)$ process for $N_\text{tot}=100$ channels, and nullclines for the deterministic $v$ (\rot{thick grey line}) and $n$ (\rot{thick black line}) dynamics. (B) Low-lying spectrum for $\LL^\dagger$ for two different channel numbers, $N_\text{tot}=100$ (black \rot{dots}) and $N_\text{tot}=25$ (red \rot{crosses}).  Dashed boxes indicate the leading complex conjugate eigenvalue pairs. (C)  Level curves (isochrons) of the asymptotic phase for $N_\text{tot}=25$ (red), $N_\text{tot}=100$ (black), and $N_\text{tot}=\infty$ (blue; deterministic case). 
The thick lines indicate the locations of the phase jump by $2\pi$, which have  been adjusted to coincide for the three cases.  Isochrons are marked in equal increments of $2\pi/20$.  Nullclines as in (A).} 
  \label{fig:napk}
\end{figure}
\end{widetext}
% PJT note (do not delete) figure from projects/aplysia/code/nap+k/matlab/napk_isochrons.m
\twocolumngrid

%The right eigenvector associated with eigenvalue $\lambda=\mu+i\omega$ for the matrix approximating $\mathcal{L}^\dagger$ approximates the eigenfunction $Q_\lambda$ 

\bi{napk}C shows level curves of the asymptotic phase function $\psi(v,n)$ in three cases, along with the nullclines from panel A.  For $N_\text{tot}\to\infty$ the process converges to the solution of a system of nonlinear ordinary differential equations for $v$ and $n$ \cite{PakdamanThieullenWainrib2010AdvAppProb}.  This system possesses a stable limit cycle for which the phase $\theta$ and isochrons are obtained in the standard way \cite{Izhikevich2007} (blue curves). Near the unstable spiral fixed point at the intersection of the nullclines, the deterministic isochrons exhibit a pronounced twisting.   For $N_\text{tot}=100$, with moderately noisy dynamics,
the level curves of the asymptotic phase $\psi$ for the stochastic system (black curves) lie close to the deterministic isochrons.  The greatest differences appear in a rarely visited region, in the neighborhood of the unstable fixed point.  As in the heteroclinic system (\bi{hetero}D), the less noisy system has more tightly wound isochrons.  For $N_\text{tot}=25$,  corresponding to an even larger noise level, the stochastic isochrons (red curves) show even less twisting. At both noise levels, the stochastic isochrons show greatest similarity to the deterministic isochrons in the region corresponding to the upstroke of the action potential, and show the greatest discrepancy at subthreshold voltages.

{\sl Discussion.}
Most investigations have approached noisy oscillators by studying the effects of weak noise on a deterministically defined phase \cite{CallenbachHaenggiLinzFreundSchimanskyGeier2002PRE,FreundSchimanskyGeierHaenggi2003Chaos,ErmentroutGalanUrban2007PRL,GalanErmentroutUrban2005PRL,ErmentroutBeverlinTroyerNetoff:2011:JCNS}.
\rot{We generalize the classical asymptotic phase to the stochastic case in terms of the eigenfunctions of the backward operator describing the evolution of densities with respect to the initial time.
As with the stochastic phase defined \textit{via} the MFPT \cite{SchwabedalPikovsky2010PRE,SchwabedalPikovsky2010EPJ,SchwabedalPikovsky2013PRL}, the backward-looking asymptotic phase is well defined whether or not the underlying deterministic system has a well defined phase.}  However, if the \rot{classical phase exists, in the absence of noise}, our  asymptotic phase is consistent with the classical definition.  

\rot{The MFPT approach has been applied to non-Markovian systems \cite{SchwabedalPikovsky2013PRL}. Our operator approach would not apply to a non-Markovian process unless it can be embedded in a higher-dimensional Markovian system \cite{HanggiJung1995AdvChemPhys}. 
Moreover, for a Markovian system, the MFPT from a point $\mbx$ to a given surface obeys an inhomogeneous partial differential equation involving the same adjoint operator $\LL^\dagger_\mbx$, an eigenfunction of which defines our asymptotic phase.   Thus,} the relationship between Schwabedal and Pikovsky's phase description of stochastic oscillators and our asymptotic phase remains an appealing topic for future research.

{\sl Acknowledgments.} PJT was supported by grant \#259837 from the Simons Foundation, by the Council for the International Exchange of Scholars, and by National Science Foundation grant DMS-1413770. BL was supported by the Bundesministerium f\"{u}r Bildung und Forschung (FKZ: 01GQ1001A).
The authors thank 
H.~Chiel, 
M.~Gyllenberg, 
L.~van Hemmen, 
A.~Pikovsky, 
M.~Rosenblum, 
L.~Schimansky-Geier, 
J.~Schwabedal,
and F.~Wolf 
for helpful discussions.
\bibliographystyle{unsrt}
%\bibliography{PJT,physics,neuroscience,math,reliability}
\bibliography{phase-corrected-2015-01-18}

\begin{thebibliography}{10}

\bibitem{ErmentroutTerman2010book}
G.B. Ermentrout and D.H. Terman.
\newblock {\em Foundations Of Mathematical Neuroscience}.
\newblock Springer, 2010.

\bibitem{Ijspeert:2008:NeuralNet}
A.J. Ijspeert.
\newblock {\em Neural Netw.}, 21:642, 2008.

\bibitem{Jordan+Smith:NODE-book:2007}
D.W. Jordan and P.~Smith.
\newblock {\em Nonlinear Ordinary Differential Equations}.
\newblock Oxford University Press, 4th edition, 2007.

\bibitem{ErmentroutKopell1984SIAMJMathAnal}
G.B. Ermentrout and N.~Kopell.
\newblock {\em SIAM J. Math Anal}, 15:215, 1984.

\bibitem{PikovskyRosenblumKurths2001}
A.~Pikovsky, M.~Rosenblum, and J.~Kurths.
\newblock {\em Synchronization: {A} universal concept in nonlinear sciences}.
\newblock Cambridge University Press, 2001.

\bibitem{noisy-oscillator-refs}
Examples include spontaneous oscillations of hair bundles in inner ear organs
  [P.~Martin, D.~ Bozovic, Y.~Choe, and A.~J.~Hudspeth. {\em J.~Neurosci.},
  23(11):4533, 2003.], stochastic oscillations of the intracellular calcium
  concentration [U.~ Kummer \textit{et al}. {\em Biophys.~J.}, 89:2005], and
  subthreshold membrane oscillations [D.~Schmitz, T.~Gloveli, J.~Behr,
  T.~Dugladze, and U.~Heinemann. {\em Neurosci.}, 85(4):999, 1998; J.A.~White,
  R.~Klink, A.~Alonso, A.R.~Kay. {\em J.~Neurophys.}, 80:262, 1998.].

\bibitem{SchwabedalPikovsky2010PRE}
J.T.C. Schwabedal and A.~Pikovsky.
\newblock {\em Phys Rev E}, 81:046218, 2010.

\bibitem{SchwabedalPikovsky2010EPJ}
J.T.C. Schwabedal and A.~Pikovsky.
\newblock {\em Eur. Phys. J.}, 187:63, 2010.

\bibitem{SchwabedalPikovsky2013PRL}
J.T.C. Schwabedal and A.~Pikovsky.
\newblock {\em Phys. Rev. Lett.}, 110:4102, 2013.

\bibitem{ShawParkChielThomas2012SIADS}
K.M. Shaw, Y-M. Park, H.J. Chiel, and P.J. Thomas.
\newblock {\em SIAM J. Appl. Dyn. Sys.}, 11:350, 2012.

\bibitem{comparison}
\rot{In the plot we omit points around $X_1=X_2=0$ (for which a reliable
  estimation of the phase was difficult) and added a small off-set to the
  remaining points for better visibility. Relative numerical error between
  theory and simulations is below 5\% for both noise levels.}

\bibitem{Izhikevich2007}
E.M. Izhikevich.
\newblock {\em Dynamical Systems in Neuroscience}.
\newblock Computational Neuroscience. MIT Press, Cambridge, Massachusetts,
  2007.

\bibitem{MorrisLecar:1981}
C.~Morris and H.~Lecar.
\newblock {\em Biophys. J.}, 35:193, 1981.

\bibitem{Rinzel+Ermentrout:1989}
J.~Rinzel and G.B. Ermentrout.
\newblock In C.~Koch and I.~Segev, editors, {\em Methods in Neuronal Modeling}.
  MIT Press, second edition, 1989.

\bibitem{details}
Applied current $I_0=60\mu\text{A/cm}^2$, passive leak current
  $I_\text{L}=g_\text{L}(V-V_\text{L})$ with $g_\text{L}=1\text{mS/cm}^2$ and
  $V_\text{L}=-78$mV, ``persistent sodium" current
  $I_\text{NaP}=\bar{g}_\text{NaP}m_\infty(V)(V-V_\text{NaP})$ with
  $\bar{g}_\text{NaP}=4\text{mS/cm}^2$, $V_{\text{NaP}}=60$mV and voltage
  dependent activation $m_\infty(v)=1/(1+\exp((-30-v)/7)))$; potassium current
  $I_\text{K}=(\bar{g}_\text{K}N/N_\text{tot})(V-V_\text{K})$ with
  $\bar{g}_\text{K}=4\text{mS/cm}^2$, $V_\text{K}=-90$mV, and open channel
  number $0\le N\le N_\text{tot}$. Membrane capacitance is
  $C=1\mu\text{F/cm}^2$. Per capita transition rate for channel opening is
  $\alpha(v)=1/(1+\exp((-45-v)/5))$ and for closing is $\beta(v)=1-\alpha(v)$.

\bibitem{AndersonErmentroutThomas2014JCNS}
D.F. Anderson, B.~Ermentrout, and P.J. Thomas.
\newblock {\em J. Comput. Neurosci.}, 38(1):67--82, 2015.

\bibitem{NewbyBressloffKeener2013PRL}
J.M. Newby, P.C. Bressloff, and J.P. Keener.
\newblock {\em Phys. Rev. Lett.}, 111:128101, 2013.

\bibitem{PakdamanThieullenWainrib2010AdvAppProb}
K.~Pakdaman, M.~Thieullen, and G.~Wainrib.
\newblock {\em Adv. Appl. Prob.}, 42:761, 2010.

\bibitem{CallenbachHaenggiLinzFreundSchimanskyGeier2002PRE}
L.~Callenbach, P.~H{\"a}nggi, S.J. Linz, J.A. Freund, and L.~Schimansky-Geier.
\newblock {\em Phys Rev E}, 65:051110, 2002.

\bibitem{FreundSchimanskyGeierHaenggi2003Chaos}
J.A. Freund, L.Schimansky-Geier, and P.~H{\"a}nggi.
\newblock {\em Chaos}, 13:225, 2003.

\bibitem{ErmentroutGalanUrban2007PRL}
G.B. Ermentrout, R.F. Gal{\'a}n, and N.N. Urban.
\newblock {\em Phys. Rev. Lett.}, 99:248103, 2007.

\bibitem{GalanErmentroutUrban2005PRL}
R.F. Gal{\'a}n, G.B. Ermentrout, and N.N. Urban.
\newblock {\em Phys. Rev. Lett.}, 94:158101, 2005.

\bibitem{ErmentroutBeverlinTroyerNetoff:2011:JCNS}
G.B. Ermentrout, B.~Beverlin, T.~Troyer, and T.I. Netoff.
\newblock {\em J. Comput. Neurosci.}, 31:185, 2011.

\bibitem{HanggiJung1995AdvChemPhys}
P.~H{\"a}nggi and P.~Jung.
\newblock {\em Adv. Chem. Phys.}, 89:239, 1995.

\end{thebibliography}
\end{document}